# Di-lepton $t\bar{t}H$ At The CMS

Sue Ann Koay

*Physics Department, University of California, Santa Barbara, California, 93106, USA*

**Abstract.** In keeping with the "find the Higgs" bandwagon, due to embark together with the Large Hadron Collider (LHC), we investigate discovery prospects with Higgs produced in conjunction with two top quarks decaying in the di-lepton channel. The following is a brief account of adventures along the way; the interested reader may find more in our CMS note [1].

**Keywords:** Higgs, $t\bar{t}H, H \to b\bar{b}$

**PACS:** 14.80.Bn.

## MOTIVATION

As the Standard Model has it, a universe with Higgs boson of mass in a range up to about 135 GeV/c$^2$ will see it decay predominantly into a pair of bottom quarks. As direct production in this channel is overwhelmed by the QCD cross-section for $b\bar{b}$ production and the inability to reconstruct the Higgs mass very precisely, one must look to modes with lower backgrounds, such as Higgs in conjunction with a $t\bar{t}$ or $b\bar{b}$ pair.

In this study, we concern ourselves with $t\bar{t}H$ production. The top quark is known to decay to a W boson and a bottom quark with a branching fraction of nearly unity; as such, $t\bar{t}H, H \to b\bar{b}$ events almost always contain four bottom quarks in the final state. Further selectivity is obtained by looking only at events with leptonic decay of both W bosons — the di-lepton channel. Although this comprises only about 5% of all signal events, it has the appeal of a rather dramatic signature: four bottom quarks, two oppositely charged leptons, and missing energy from two neutrinos.

## STRATEGY

### Event Generation

The date of writing must excuse the fact that this is a study of Monte Carlo data. The well-established leading order Monte Carlo, PYTHIA, is used to generate signal ($t\bar{t}H$) samples with at least one W boson decaying leptonically. In order to take into account that the most dangerous backgrounds are top pairs produced in conjunction with some number of hard jets, the ALPGEN[1] [2] program is used to perform calculations using higher order matrix elements that include additional radiated partons. The backgrounds included in our study are $t\bar{t}$ with up to four light-quark jets, $t\bar{t}b\bar{b}$, and $t\bar{t}Z$. Both signal and background and subjected to a full simulation of the Compact Muon Solenoid (CMS) detector, with every attempt made to be as realistic as possible within current knowledge of how an analysis of actual data would carry out.

---

[1] Interfaced to PYTHIA for parton showering.

## Event Selection

The presence of two high-energy, oppositely charged leptons is of course much of the charm of the di-lepton channel. For selecting the leptons from the signal event, we use a likelihood method designed to distinguish them from narrow jets and also other real leptons in background events. The cuts are chosen to correspond to an 85% efficiency for selecting actual signal leptons, versus a 3% chance for mistakes.

Missing transverse energy is present in most di-lepton events, since the two signature neutrinos do not register in any part of the detector. The "raw" missing energy computed from all calorimeter tower energies is corrected to account for missing energy carried by muons as well as under-reported energies of jets. We then require the value to be at least 40 GeV.

Jets of interest in di-lepton $t\bar{t}H$ events should in principle all be b-tagged. The CMS Combined Secondary Vertex b-tagging algorithm computes a discriminator variable for jets, categorized by whether or not a secondary vertex is reconstructed. To illustrate, variables for the "has vertex" category include the impact parameter significances for tracks and also for the secondary vertex, the vertex mass, etc. A likelihood method similar to that for lepton identification is applied to obtain the discriminator. The working point used in our analysis corresponds to a roughly 60% bottom-jet tagging efficiency, at the cost of < 20% charm and 3% collectively of up, down, strange, and jets from gluon splitting.

With four b-jets from the signal event and more due to the high center-of-mass energy at the LHC, jet calibration plays a significant role in event selection and eventual (one hopes) reconstruction of the Higgs. Several calibration schemes have been investigated, but seem on the whole to be of comparable performance. In Figure 1 we see that — if we are able to correctly guess which two bottom quarks came from the Higgs decay (nontrivial!) — we can expect to reconstruct the Higgs mass to 20% or so.

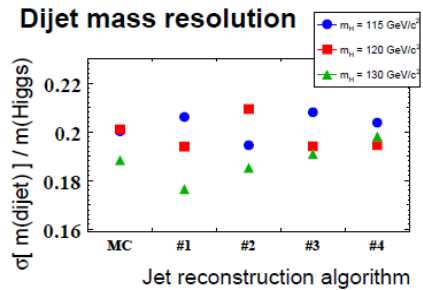

**FIGURE 1.** "Best-case" Higgs mass resolution for various jet reconstruction algorithms.

The combinatorics problem in assigning jets to the four bottom quarks is not easily resolved for the di-lepton channel, because the two neutrinos present too many degrees of freedom for, say, reconstructing the top quarks. In the current counting experiment, we have not made the effort, and indeed had hopes of avoiding it. Thus, in the jet selection department, we "merely" require four to seven jets that are within tracker acceptance and have transverse energy greater than 20 GeV, three or four of which must be b-tagged.

## PROCEEDS

Before systematics are taken into account, signal significances ranging from about 1.8 ($m_H$ = 115 GeV/c$^2$) to 0.86 ($m_H$ = 130 GeV/c$^2$) are predicted for our simple counting experiment at 60 fb$^{-1}$, as shown in Figure 2. By far the most significant background, $t\bar{t}$ +jets also includes an irreducible $t\bar{t}b\bar{b}$ piece. Both require additional ingenuity to tame if we are to hope to observe a signal in this Higgs decay channel. Other conceivable backgrounds, such as W+jets, Z+jets, and di-boson+jets, have been estimated by way of simple calculations to be negligible in comparison.

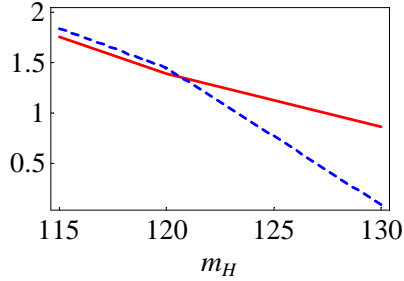

**FIGURE 2.** Signal significance at 60 fb$^{-1}$ as a function of Higgs mass, for the di-lepton channel only.

With systematics properly included, the signal significance inevitably drops. The most dangerous systematic of those considered for this analysis is that associated with the tagging efficiency for light-quark jets, although the jet energy scale and jet resolution uncertainties are close contenders. The total uncertainty in the background, summed by quadratures as appropriate for independent systematics, comes out to be around 18% of the background.

Sadly, even in combination with the other $t\bar{t}H, H \to b\bar{b}$ channels, i.e. with semi-leptonic and all-hadronic decays of the $W$ bosons, the signal significance falls rapidly to zero as a function of the total systematic uncertainty relative to the total background. Tighter cuts than those sketched above may help, and there are several other techniques that can be applied to data that were not relevant for Monte Carlo. However, even in the best scenarios, observing the Higgs in this channel looks to be disappointing. A more thorough discussion can be found in the CMS note [1].